\documentclass[aps,prl,twocolumn,groupedaddress]{revtex4-1}

\usepackage{graphicx}
\usepackage{amsmath}
\usepackage{color}

\begin{document}

%\preprint{}

\title{Fermi surface deformations and pairing in mixtures of dipolar and non-dipolar fermions.}


\author{J.E. Baarsma}
\author{P. T{\"o}rm{\"a}}
\email[]{paivi.torma@aalto.fi}
%\thanks{}
\affiliation{COMP Centre of Excellence, Department of Applied Physics, Aalto University, FI-00076 Aalto, Finland}
%\affiliation{ITP Utrecht, The Netherlands}


\date{\today}

\begin{abstract}
We study mass-imbalanced two-component Fermi mixtures, where one of the components consists of dipolar fermions. We specifically study the mass imbalances corresponding to the atomic ${}^{163}$Dy-${}^{40}$K and ${}^{53}$Cr-${}^{6}$Li mixtures. We study the onset of the $s$-wave superfluid phase, as a function of population imbalance and the dipolar interaction strength. We find the critical temperature and the Fermi surface deformations at the transition to depend on the dipolar interaction strength, where the critical temperature increases (decreases) with dipolar interaction strength for a majority (minority) of dipolar atoms. We present momentum distributions of both components where the Fermi surface deformations are visible. 
\end{abstract}

%\keywords{}
\maketitle
%\section{}
%\subsection{}
%\subsubsection{}

\emph{Introduction.} -- Anisotropic, long-range interactions, combined with ultralow temperatures are predicted to lead to a whole range of exotic physical phenomena, such as supersolidity \cite{goral_quantum_2002,yi_novel_2007,he_supersolid_2011} and topological superfluids \cite{cooper_stable_2009,levinsen_topological_2011}, as well as unconventional magnetic states \cite{fregoso_ferronematic_2009,fregoso_unconventional_2010}. 
In the field of ultracold dipolar Bose gases, the progress is huge, both in experimental and theoretical research \cite{baranov_theoretical_2008,lahaye_physics_2009,baranov_condensed_2012,torma_dipolar_2014,torma2_dipolar_2014}, examples of which are the observation of an anisotropic deformation during expansion of an atomic cloud, the tuning of the dipolar interaction between dipolar molecules and the observation of interaction-induced crystallization in a superfluid \cite{krems_cold_2009,lu_quantum_2012,kadau_observing_2016}.
%The extremely high controllability in experiments using ultracold Fermi gases has lead already to many amazing discoveries \cite{bloch_many-body_2008,chevy_ultra-cold_2010,gubbels_imbalanced_2013}. Especially the use of Feshbach resonances to tune the isotropic $s$-wave interaction resulted in a number of interesting experiments, of which the discovery of the BEC-BCS crossover was ground breaking \cite{regal_observation_2004}. Examples of more recent exciting developments are the realization of the Fermi microscope \cite{cheuk_quantum-gas_2015,parsons_site-resolved_2015,haller_single-atom_2015,edge_imaging_2015,omran_microscopic_2015} and the study of the 1D-3D crossover \cite{revelle_1d_2016}.
Also in dipolar Fermi gases exotic states of matter are expected. For instance, the attractive part of the dipolar interaction can lead to a $p$-wave superfluid, which was not yet realized experimentally, but has been studied extensively in the literature \cite{stoof_condensed_1998,baranov_superfluid_2002,baranov_superfluidity_2004,bruun_quantum_2008,sun_spontaneous_2010,gadsbolle_dipolar_2012,corro_effect_2016,fedorov_novel_2016}. For strong dipolar interactions the Fermi gas collapses, while the dipolar gas is in a metastable state for small interactions due to the Fermi pressure \cite{miyakawa_phase-space_2008,bruun_quantum_2008,ronen_zero_2010,parish_density_2012}. In a two-dimensional lattice, liquid crystal phases and complicated density-ordered phases have been predicted \cite{lin_liquid_2010,mikelsons_density-wave_2011}.

Experimentally, fermionic dipolar molecules in their rovibrational and hyperfine ground state were created \cite{ni_high_2008,park_ultracold_2015,moses_creation_2015} and quantum degenaracy has been reached in Fermi gases of dipolar atoms \cite{lu_quantum_2012,aikawa_reaching_2014,naylor_chromium_2015}. 
Recent experiments have shown that when the dipole moments of the atoms are aligned by an external field, the anisotropic dipolar interaction leads to real-space nematic ordering of the gas \cite{aikawa_anisotropic_2014,aikawa_observation_2014}. This shows up in momentum space as an elliptical Fermi surface, where the elongation can be controlled by adjusting the external field \cite{aikawa_observation_2014}. The latter has been studied before in a number of theory works \cite{miyakawa_phase-space_2008,sogo_dynamical_2009,ronen_zero_2010,krieg_second-order_2015}. 

\begin{figure*}\begin{center}
\includegraphics[width=.45\textwidth]{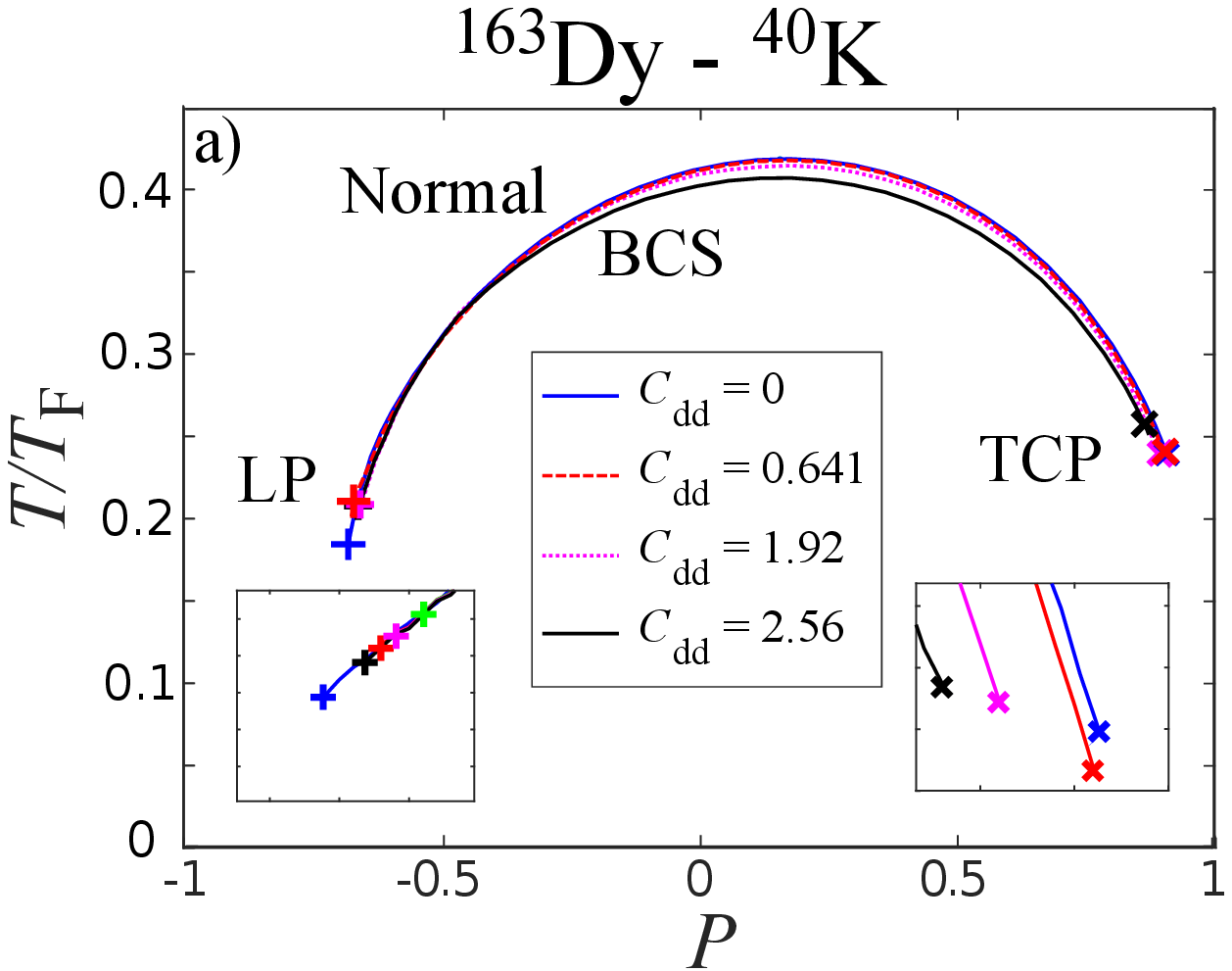}
\includegraphics[width=.45\textwidth]{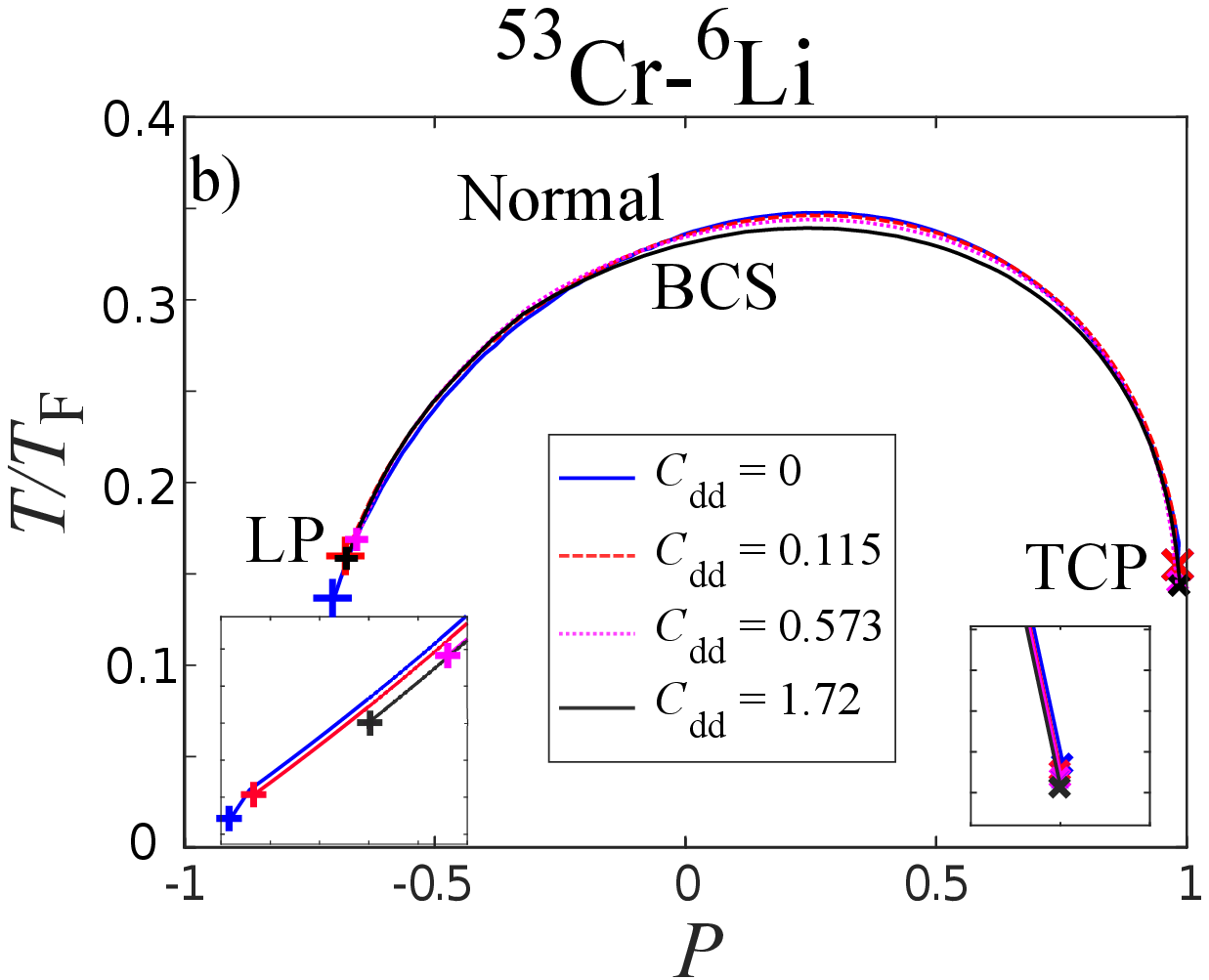}
\end{center}
\caption{(Color online) Phase diagrams for examples of Fermi mixtures consisting of dipolar and non-dipolar particles, namely the Dy-K mixture (left) and the Cr-Li mixture (right) for different dipolar interaction strengths $C_\text{dd}$. The lines denote continuous phase transitions from a normal state to a superfluid (BCS) and $C_\text{dd}$ increases from the top to the bottom for $P=0$. For a majority of dipolar atoms, $P<0$, a Lifshitz point (LP) is present in the phase diagram and a tricritical point (TCP) for $P>0$. The insets are enlargements of the LP and TCP regions.}
\label{ddresults}
\end{figure*}

\begin{figure*}
\begin{center}
\includegraphics[width=.97\textwidth]{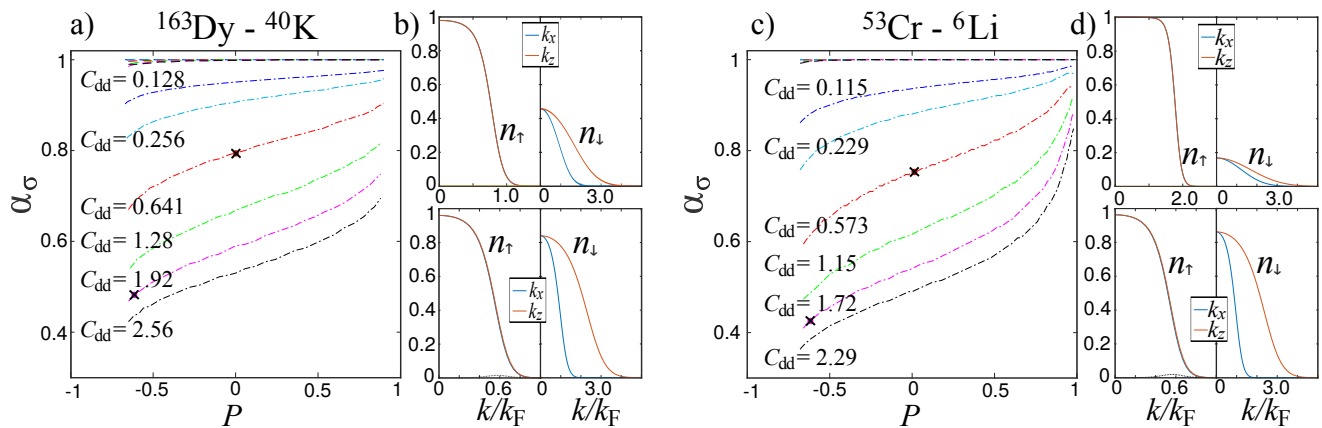}
\end{center}
\caption{(Color online) Deformations $\alpha_\sigma$, a) and c), along the lines of the phase transitions in Fig. 1. The momentum distributions for both components are presented along both the $k_x$ and the $k_z$ directions in b) and d) for the marked points in a) and c), respectively. In the upper panels $P=0$ and in the lower $P=-0.62$. For the $\uparrow$ component also the difference is shown.}
\label{defmom}
\end{figure*}

In this Letter we focus on systems combining both interspecies dipolar and intraspecies $s$-wave interactions, which, as far as we are aware of, have not been considered before. We study mixtures of dipolar and non-dipolar fermionic atoms, where we assume the dipole moments of the dipolar component to all be aligned. Consequently, we study the pairing instability between the two components within a BCS mean-field theory. We investigate the influence of the dipolar interaction on the onset of superfluidity and in turn how the $s$-wave interaction affects the nematic ordering of the dipolar component. After elaborating on the methods used, results are presented for the mass-imbalances corresponding to the experimentally available ${}^{163}$Dy-${}^{40}$K and ${}^{53}$Cr-${}^{6}$Li mixtures. We find an instability towards FFLO phases in both mixtures for a majority of heavy, dipolar atoms, where the critical temperature and polarization of the Lifshitz point depend on the dipolar interaction strength, see Fig.\ref{ddresults}. This is an interesting prospect for the possible observation of the elusive FFLO states. We find that the scaling of the critical temperature $T_c$ with dipolar interaction strength very much depends on whether the dipolar atoms form the majority or the minority component. In the first case $T_c$ increases for stronger interactions, while in the latter it decreases.
Also, we find that the Fermi surface deformation of the dipolar component depends strongly on the presence of the non-dipolar component, see Fig.\ref{defmom}, and the $s$-wave interaction induces the anisotropic character of the dipolar interaction to be reflected also in the non-dipolar component. We show that the anisotropy is visible in the momentum distributions of both components, which can be measured in experiment by a time of flight measurement.

\emph{Methods.}---
In this work, we use a BCS mean-field theory, where we incorporate the possibility of elliptic Fermi surfaces by taking both particle dispersions to be anisotropic. We arrive at a total energy functional $E_\text{tot}$ by adding the dipolar interaction energy to the BCS thermodynamic potential, which reads
\begin{align}
\nonumber \omega_\text{BCS}=&-\frac{|\Delta|^2}{T^{2\text B}}\int\frac{\text{d}{\bf k}}{(2\pi)^3}\Big\{  \varepsilon_{\bf k}^\alpha-\mu-\hbar\omega_{\bf k}^\alpha\\
&+\frac{\Delta^2}{2\varepsilon_{\bf k}^\alpha}
-\frac1\beta\sum_{\sigma=\uparrow,\downarrow}\log\left[1+e^{-\beta\hbar\omega_{\sigma{\bf k}}}\right]\Big\}.
\label{ombcs}
\end{align} 
The two-body scattering matrix is related to the $s$-wave scattering length $a$ via $T^{2\text B}= m/(4\pi\hbar^2a)$, with $m$ twice the reduced mass, $m=2m_\uparrow m_\downarrow/(m_\uparrow+m_\downarrow)$. In this Letter, we consider the unitarity limit, where $1/(k_Fa)=0$, and since we only consider the regime where the $s$-wave interaction is much stronger than the dipolar interaction, we do not consider $p$-wave pairing between the dipolar atoms. 

Here, the Cooper pairs, described by $\Delta$ in $\omega_\text{BCS}$, consist of one non-dipolar ($\downarrow$) and one dipolar ($\uparrow$) atom. The average dispersion in Eq.(\ref{ombcs}) is $\varepsilon_{\bf k}^\alpha-\mu=(\xi_\uparrow+\xi_\downarrow)/2$, where we take the possibility of elongated Fermi surfaces into account by using anisotropic dispersion relations,
\begin{align}
\xi_{\sigma{\bf k}}=\frac{\hbar^2}{2m_\sigma}\left(\frac{k_\perp^2}{\alpha_\sigma}+\alpha_\sigma^2k_z^2 \right)-\mu_\sigma,
\label{disp}
\end{align}
with $\mu_\sigma$ the chemical potential of the $\sigma$ component. 
The quasi-particle dispersions are $\hbar\omega_\sigma=\hbar\omega\pm(\xi_\uparrow+\xi_\downarrow)/2$, with the +(-) corresponding to the $\uparrow(\downarrow)$ quasi-particles and $\hbar\omega=\sqrt{[\varepsilon_{\bf k}^\alpha-\mu]^2+\Delta^2}$. 

In all equations, the superscript $\alpha$ denotes that the corresponding term depends on both parameters $\alpha_\sigma$, which define the deformations of the Fermi surfaces. Namely, the dispersions are isotropic for $\alpha_\sigma=1$, while the Fermi momenta $k_\text{F}$ in the $k_\perp$ and $k_z$ directions are different whenever $\alpha_\sigma$ differs from one. For example, at zero temperature $k_{\text{F},z}=k_\text{F}^0/\alpha_\sigma$ and $k_{\text{F},\perp}=\sqrt{\alpha_\sigma}k_\text{F}^0$ for a non-interacting system, where $k_\text{F}^0$ denotes the Fermi momentum in the isotropic case. For a dipolar gas with aligned dipole moments, the particles arrange in such a way that $\alpha<1$, which thus means $k_{F,z}>k_{F,\perp}$ \cite{aikawa_observation_2014}.
We choose this specific dependence of the dispersions on the deformation $\alpha_\sigma$, because we assume the dipoles to be aligned in the $z$-direction and in order for the non-interacting particle densities $n_\sigma$ to be invariant under deformations, $n_{\sigma}^\alpha=n_{\sigma}^0$ \cite{miyakawa_phase-space_2008,sogo_dynamical_2009}.

It is ensured that the energy of the system in the absence of dipolar interactions is minimized by isotropic Fermi surfaces, by the following term
\begin{align}
E_\text{iso}=\int\frac{\text{d}{\bf k}}{(2\pi)^3}\sum_\sigma\left[\varepsilon_{\sigma{\bf k}}^\alpha-\varepsilon_{\sigma{\bf k}}^0\right]N_{F}\left(\varepsilon_{\sigma{\bf k}}^\alpha\right),
\end{align}
where the superscript $0$ denotes that $\alpha_\uparrow=\alpha_\downarrow=1$ in that term. This energy is zero for isotropic Fermi surfaces and positive for any $\alpha_\sigma$ being different from one. In other words, $E_\text{iso}$ describes the cost in kinetic energy when deforming a Fermi surface.

Naturally, the last part of our energy functional is the interaction energy for aligned dipoles \cite{miyakawa_phase-space_2008,sogo_dynamical_2009}
\begin{align}
E_\text{dd}=-\frac{\pi n_\downarrow^{5/3}m_\downarrow C_\text{dd}}{3\hbar^2}\int_0^\pi\text{d}\theta\sin\theta\Bigg(\frac{3\cos^2\theta}{\alpha_\downarrow^3\sin^2\theta+\cos^2\theta}-1\Bigg).
\end{align}
The dimensionless dipolar interaction strength is $C_\text{dd}=m_\downarrow n_\downarrow^{1/3}d^2/\hbar^2$,
with $d$ the dipole moment, which is $d=9\mu_\text{B}$ and $d=6\mu_\text{B}$, with $\mu_\text{B}$ the Bohr magneton, for Dy and Cr respectively. The attainable values for $C_\text{dd}$ depend on the densities $n_\downarrow$ and it is, for example, $C_\text{dd}\approx 0.2$  for a homogeneous Dy gas with a density of $10^{14}$ cm$^{-3}$. In a trapped system the Fermi surface deformations also depend on the trapping frequencies and we consider values for $C_\text{dd}$ corresponding to experimentally attained deformations \cite{aikawa_observation_2014} .

In this Letter, we determine the phase transitions by minimizing $E_\text{tot}=\omega_\text{BCS}+E_\text{iso}+E_\text{dd}$ with respect to $\Delta$ and both $\alpha_\sigma$ at different temperatures and polarizations, where the polarization $P=(n_\uparrow-n_\downarrow )/n$, with $n$ the total density. The gap $\Delta$ is the order parameter for the phase transition from a normal Fermi gas to a BCS superfluid of condensed Cooper pairs. In the normal state, $\Delta=0$, both deformations can be scaled out from $\omega_\text{BCS}$, whereas the thermodynamic potential only depends on the ratio $\alpha_\uparrow/\alpha_\downarrow$ in the case of a non-zero gap. The dipolar interaction energy is minimized by $\alpha_\downarrow=0$ and the competition between the different terms in $E_\text{tot}$ determines $\Delta$ and $\alpha_\sigma$. Here, we study the $s$-wave pairing instability and determine the Fermi surface deformations right at the onset of superfluidity.

We find for both Fermi mixtures under study continuous transitions from a normal gas to a homogeneous superfluid phase. These transition lines end for both mixtures at a tricritical point end a Lifshitz point, depending on polarization. At a tricritical point the character of the phase transition turns from continuous into a discontinuous transition, where the latter is associated with phase separation. The tricritical point can be found by studying whether the global minimum of $E_\text{tot}$ at a nonzero $\Delta$ develops continuously or discontinuously.
The Lifshitz point is a multicritical point, at which the homogenous superfluid phase, the normal phase and an FFLO region meet. Below the Lifshitz point the pairing instability is towards a superfluid phase where Cooper pairs carry a net momentum, as a nonzero kinetic energy for the Cooper pairs is energetically favorable inside the FFLO phase. We find the Lifshitz points by including also the Cooper pair propagator in our theory and determining when the effective mass of the Cooper pairs is zero \cite{gubbels_lifshitz_2009,baarsma_population_2010}.

\emph{Phase Diagrams.} --- The critical temperatures for the transitions depend on the population imbalance and the dipolar interaction strength.
In Fig.\ref{ddresults} the phase diagrams as a function of polarization $P$ and temperature $T$ are presented for the ${}^{163}$Dy-${}^{40}$K (left) and the ${}^{53}$Cr-${}^{6}$Li (right) mixtures. 
The critical temperatures are scaled by the Fermi temperature $k_\text{B}T_\text{F}=\hbar^2(3\pi n)^{2/3}/(2 m)$. The transition lines correspond to continuous transitions and are shown for different values of $C_\text{dd}$.

We find in the phase diagrams of both the Dy-K and the Cr-Li mixtures a Lifshitz point for a majority of dipolar atoms, and a tricritical point for a majority of non-dipolar atoms, which is similar to the phase diagram of the non-dipolar ${}^6$Li-${}^{40}$K mixture \cite{gubbels_lifshitz_2009,baarsma_inhomogeneous_2013}. The occurrence of both these multicritical points results from the mass and population imbalance in the Fermi mixtures, namely, we find both also in the absence of the dipolar interaction, $C_\text{dd}=0$, see Fig.\ref{ddresults}.

For stronger dipolar interactions $C_\text{dd}$, the deformations of the Fermi surfaces are larger and the critical temperature is different, where, interestingly, the shift of the critical temperature depends on the polarization in the mixture. For equal densities and for a majority of non-dipolar particles, the critical temperature decreases with increasing dipolar interaction strength. However, if the dipolar atoms form the majority component, the critical temperature shift with this scaling is much smaller and is actually an increase. We study these effects in more detail in Fig. 3. 
The locations of the multicritical points in the phase diagrams are also affected by the dipolar interaction. 
%Namely, the tricritical point at first shifts towards higher temperature and smaller polarization with increasing dipolar interaction, but then shifts again down in temperature. The Lifshitz points first occur at smaller polarization and higher temperature with increasing dipole interactions, but then also shift towards lower temperatures again.

Determining the first-order transition lines and the size and exact structure of the FFLO region is beyond the scope of this Letter, since for both it is not enough to study the onset of superfluidity only. Instead, the effect of the dipolar interactions and the Fermi surface deformations in the superfluid phase should then be incorporated. The transition lines in Fig.\ref{ddresults} therefore end at the multicritical points.

\emph{Fermi surface deformations.} --- 
The two deformation parameters $\alpha_\sigma$ at the phase transition found by minimizing $E_\text{tot}$ are shown in Fig.\ref{defmom}, and an interesting effect that can be observed is that the deformation $\alpha_\downarrow$ here not only depends on the dipolar interaction strength and the temperature, but also largely on the presence of the non-dipolar component. Namely, along each transition line we kept both $C_\text{dd}$ and the density of the dipolar atoms $n_\downarrow$ fixed, in order for the dipolar interaction energy to be constant. 
Moreover, the different scaling of $\alpha_\downarrow$ with $P$ than the temperature demonstrates that it is not solely a temperature effect, but rather is an effect of the $s$-wave interactions.

Interestingly, for $P\ll 0$ also the non-dipolar cloud starts to deform and also $\alpha_\uparrow< 1$, although the deformation is much smaller. 
For the mixture with the larger mass imbalance, Cr-Li, the dipolar component deforms more strongly compared to the Dy-K mixture, whereas the non-dipolar component deforms more in the latter mixture.

We also considered mass balanced systems, where the phase diagram is symmetrical in $P$ without dipolar interactions and only contains tricritical points. We found $T_c$ to change differently with $C_\text{dd}$ for $P>0$ and $P<0$ and we found for both components a larger deformation $\alpha_\sigma$ compared to the mass-imbalanced mixtures.

The parameters $\alpha_\sigma$ differing from 1 results in a different shape of the momentum distributions in different directions, see Fig.\ref{defmom}b) and d). We calculated the Fermi momentum distributions for the points marked in Fig.\ref{defmom} a) and c) and indeed the distributions decay to zero at different momenta in the $k_z$ and $k_x$ direction. For the $\uparrow$ component also the difference is shown.

These momentum distributions could be measured, when the clouds expand ballistically, in a time-of-flight experiment and the deformation effects are big enough to observe. We did not take into account the non-ballistic expansion for the dipolar component, which leads to quantitative differences in the measured momentum distributions \cite{sogo_dynamical_2009,veljic_time--flight_2016}.

\begin{figure}\begin{center}
\includegraphics[width=\columnwidth]{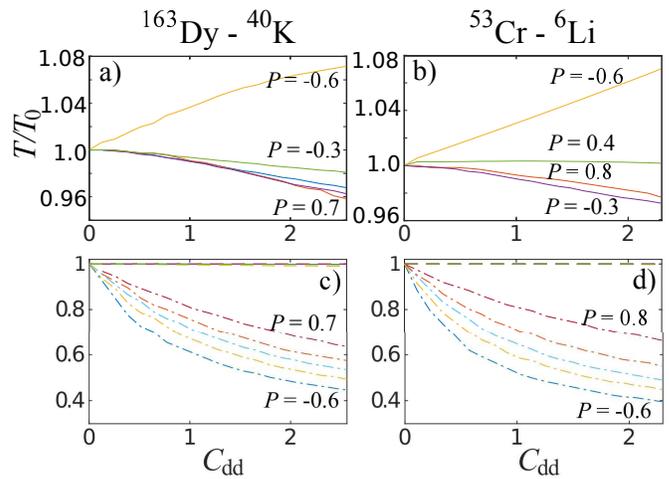}
\end{center}
\caption{(Color online) Critical temperature $T_c$, scaled by the critical temperature $T_0$ for $C_\text{dd}=0$, a) and b), and deformations $\alpha_\sigma$, c) and d), as a function of $C_\text{dd}$ for different polarizations $P$. In c) and d) the dashed (dashed-dotted) lines correspond to the $\uparrow$($\downarrow$) particles and $P$ decreases from the top to the bottom.}
\label{Cddresults}
\end{figure}

\emph{Effect of the dipolar interaction.} --- In Fig. \ref{Cddresults}, we determine at fixed polarizations the critical temperature $T_c$ and the deformations $\alpha_\sigma$, for the transition to the BCS phase, as a function of the dipolar interaction strength $C_\text{dd}$, in order to study in more detail the effect of the dipolar interactions on the BCS transition. Here the temperatures are scaled by $T_0$, the critical temperature for $C_\text{dd}=0$ at the corresponding polarization. Both deformations $\alpha_\uparrow$ and $\alpha_\downarrow$ decrease with increasing interaction strength. 
In contrast, the effect of the dipolar interactions on the critical temperature depends on the polarization and can be both a temperature increase or decrease, where the effects are more pronounced for a larger population imbalance.
The latter demonstrates that the deformation of the Fermi surface can have a beneficial effect on the BCS pairing. 

Interestingly, the effect of deformed Fermi surfaces at the pairing instability has been studied before in the context of imbalanced non-dipolar Fermi systems, as an alternative for FFLO phases \cite{muther_spontaneous_2002,sedrakian_pairing_2005}.

\emph{Conclusions and Discussion.} --- We studied the onset of superfluidity in mass-imbalaced Fermi mixtures consisting of dipolar and non-dipolar atoms. We presented the phase diagrams for the ${}^{163}$Dy-${}^{40}$K and the ${}^{53}$Cr-${}^{6}$Li mixtures and found that the phase diagrams contain both a tricritical point and a Lifshitz point. We found the critical temperature and the Fermi surface deformations to depend on the dipolar interaction strength.
 
This work is a first step in exploring the physics that arises from combining isotropic and anisotropic interactions in Fermi mixtures, and is a good starting point for studying these systems using beyond mean-field methods needed for better quantitative predictions \cite{PhysRevB.91.144510}. 
For the possible observation of the FFLO phase it is a very interesting prospect that the parameter range for the FFLO phase depends on $C_\text{dd}$ and to study the FFLO states in the presence of dipolar interaction would be an interesting continuation of this work.

Another interesting extension is to consider dipolar and $s$-wave interaction strengths of similar order of magnitude. The effect of different scattering lengths, $k_\text{F} a$, on the BCS transition has been studied elaborately, also in mass-imbalanced Fermi mixtures \cite{PhysRevA.75.063601,baarsma_population_2010}.

The mass ratios of ${}^{163}$Dy-${}^{40}$K and  of ${}^{53}$Cr-${}^{6}$Li, 4.075 and 8.833 respectively, are well below the ratio 13.6 above which Efimov three-body bound states exist \cite{1973NuPhA.210..157E}. However, at unitarity and for repulsive interactions, Kartavtsev-Malykh trimers, or related resonant enhancement of $p$-wave interaction between heavy-light dimers and heavy particles, may appear for these mass ratios \cite{0953-4075-40-7-011,PhysRevA.86.062703,PhysRevLett.112.075302,PhysRevA.93.053611}. Such effects could influence many-body pairing properties and the Fermi surface deformations in interesting ways: this is an important future research topic.

In this mean-field exploration we consider homogeneous systems only. The effect of a trapping potential has been shown to be important for dipolar gases \cite{aikawa_observation_2014,baranov_superfluidity_2004,sogo_dynamical_2009} and taking a trap into account would be an obvious extension of the present work, although with the recent development of box potentials also homogeneous systems can be realized experimentally \cite{meyrath_bose-einstein_2005,es_box_2010,gaunt_bose-einstein_2013,mukherjee_homogeneous_2016}.

\emph{Acknowledgements.} --- Jami Kinnunen and Henk Stoof are thanked for useful discussions. This work was supported by the Academy of Finland through its Centers of Excellence Programme (2012-2017) and under Projects No. 263347, No. 251748, No. 272490, and by the European Research Council (ERC-2013-AdG-340748-CODE).

\bibliography{DFS}
\end{document}